\pdfoutput=1

\documentclass[11pt,twoside,a4paper,cmspaper,final,collab]{cms-tdr}
\def\svnVersion{167111}\def\cmsCernNo{2012-124}\def\cmsCernDate{\today}\def\cmsMessage{Submitted to Physics Letters B}

\begin{document}\cmsNoteHeader{BPH-11-007}

\hyphenation{had-ron-i-za-tion}
\hyphenation{cal-or-i-me-ter}
\hyphenation{de-vices}

\RCS$Revision: 123752 $
\RCS$HeadURL: svn+ssh://svn.cern.ch/reps/tdr2/papers/BPH-11-007/trunk/BPH-11-007.tex $
\RCS$Id: BPH-11-007.tex 123752 2012-05-23 15:51:52Z kaulmer $
\newlength\cmsFigWidth
\ifthenelse{\boolean{cms@external}}{\setlength\cmsFigWidth{0.85\columnwidth}}{\setlength\cmsFigWidth{0.4\textwidth}}
\ifthenelse{\boolean{cms@external}}{\providecommand{\cmsLeft}{top}}{\providecommand{\cmsLeft}{left}}
\ifthenelse{\boolean{cms@external}}{\providecommand{\cmsRight}{bottom}}{\providecommand{\cmsRight}{right}}
\renewcommand{\PgLb}{\ensuremath{\Lambda_\mathrm{b}}} 
\newcommand{\PagLb}{\ensuremath{\overline{\Lambda}_\mathrm{b}}}
\providecommand{\rd}{\ensuremath{\mathrm{d}}}
\newcommand{\invfb}{\fbinv}
\newcommand{\tev}{\TeV}
\newcommand\dbline{\noalign{\vskip 0.10truecm\hrule}\noalign{\vskip 2pt}\noalign{\hrule\vskip 0.10truecm}}
\newcommand\sgline{\noalign{\vskip 0.10truecm\hrule\vskip 0.10truecm}}
\newcommand{\MLb}{\ensuremath{m_{\psilam}}}
\newcommand{\calL}{{\ensuremath{\cal L}}\xspace}
\newcommand{\calB}{\ensuremath{\mathcal{B}}}
\newcommand{\calA}{\ensuremath{\mathcal{A}}}
\newcommand{\dsdpt}{\ensuremath{\rd\sigma/\rd p_\mathrm{T}^{\PgLb}}}
\newcommand{\dsdy}{\ensuremath{\rd\sigma/\rd y^{\PgLb}}}
\newcommand{\dsdptbf}{\ensuremath{\rd\sigma/\rd p_\mathrm{T}^{\PgLb}\times\BLbpsilam}}
\newcommand{\dsdybf}{\ensuremath{\rd\sigma/\rd y^{\PgLb}\times\BLbpsilam}}
\newcommand{\aPPratio}{\ensuremath{\sigma(\PagLb)/\sigma(\PgLb)}}
\newcommand{\ptb}{\ensuremath{p_\mathrm{T}^{\PB}}}
\newcommand{\ptLb}{\ensuremath{p_\mathrm{T}^{\PgLb}}}
\newcommand{\yb}{\ensuremath{\abs{y^{\PB}}}}
\newcommand{\yLb}{\ensuremath{\abs{y^{\PgLb}}}}
\newcommand{\psimumu}{\ensuremath{\JPsi\to\Pgmp\Pgmm}}
\newcommand{\psilam}{{\ensuremath{\JPsi\, \PgL}}}
\newcommand{\lamppi}{\ensuremath{\PgL\rightarrow\Pp\Pgp}}
\newcommand{\Lbpsilam}{\ensuremath{\PgLb\to\JPsi\PgL}}
\newcommand{\BLbpsilam}{\ensuremath{\calB(\Lbpsilam)}}
\newcommand{\mev}{\ensuremath{\MeV}}
\newcommand{\pp}{\ensuremath{\Pp\Pp}}
\newcommand{\Htobb}{\ensuremath{\PHz\to\cPqb\cPaqb}}
\newcommand\T{\rule{0pt}{2.6ex}}
\newcommand\B{\rule[-1.2ex]{0pt}{0pt}}
\providecommand{\cPqb}{\ensuremath{\mathrm{b}}} 
\providecommand{\cPaqb}{\ensuremath{\overline{\mathrm{b}}}} 
\newcommand{\totalsigma}{\ensuremath{1.16\pm 0.06\pm 0.12~\textrm{nb}}}
\newcommand{\totalasym}{\ensuremath{1.02\pm 0.07\pm 0.09}}
\cmsNoteHeader{BPH-11-007} 
\title{Measurement of the \PgLb\ cross section and the \PagLb\ to \PgLb\ ratio
with $\JPsi\PgL$ decays in \pp\ collisions at $\sqrt{s} = 7\TeV$}

\date{\today}

\abstract{
The \PgLb\ differential production cross section and the cross-section ratio \aPPratio\ are measured as functions
of transverse
momentum \ptLb\ and rapidity \yLb\ in \pp\ collisions at $\sqrt{s} = 7\TeV$ using data
collected by the CMS experiment at the LHC. The measurements are based on \PgLb\ decays reconstructed
in the exclusive final state $\JPsi\Lambda$, with the subsequent decays \psimumu\ and \lamppi, using a data
sample corresponding to an integrated luminosity of $1.9\invfb$.
The product $\sigma(\PgLb)\times\BLbpsilam$ versus \ptLb\ falls faster than that of \cPqb\ mesons.
The measured value of $\sigma(\PgLb)\times\BLbpsilam$ for $\ptLb\ > 10\GeV$ and $\yLb\ < 2.0$ is
\totalsigma, and the integrated \aPPratio\ ratio is \totalasym, where the uncertainties are statistical and
systematic, respectively.}

\hypersetup{%
pdfauthor={CMS Collaboration},%
pdftitle={Measurement of the Lambda(b) cross section and the anti-Lambda(b) to Lambda(b) ratio
with Lambda b to J/Psi Lambda decays in pp collisions at sqrt(s) = 7 TeV},%
pdfsubject={CMS},%
pdfkeywords={CMS, B physics}}

\maketitle 

\section{Introduction}
Cross sections for \cPqb-quark production in high-energy hadronic collisions
have been measured at $\Pp\Pap$ colliders
at center-of-mass energies from 630\GeV~\cite{Albajar:1988th} to 1.96\TeV
~\cite{Abe:1995dv,Abulencia:2006ps,Abachi:1994kj}, in fixed-target $\Pp$-nucleus collisions with
beam energies from 800 to 920\GeV~\cite{Zaitsev:2009zz}, and recently in
$\Pp\Pp$ collisions at 7\TeV at the Large Hadron Collider
(LHC)~\cite{BPH-10-004,BPH-10-005,BPH-10-007,BPH-10-013,LHCb,Chatrchyan:2012dk,ATLAS:2011ac,Chatrchyan:2012hw}.
As the expected cross sections can
be calculated in perturbative quantum chromodynamics (QCD), the comparison between data and
predictions provides a critical test of next-to-leading-order (NLO) calculations~\cite{NDE,Cacciari04}.

Considerable progress has been achieved in understanding heavy-quark
production at Tevatron energies, largely resolving earlier discrepancies in which theoretical
predictions were significantly below observed production rates~\cite{Cacciari04}.
However, substantial theoretical uncertainties on production cross sections
remain due to the dependence on the renormalization and
factorization scales. Measurements of \cPqb-hadron production at 7\TeV\ represent
a test of theoretical approaches that aim to describe
heavy-flavor production at the new center-of-mass energy~\cite{Cacciari98,Kniehl08}.
Furthermore, understanding the production rates for
\cPqb\ hadrons represents an essential component in accurately estimating heavy-quark backgrounds
for various searches, such as \Htobb\ and supersymmetric or exotic new physics
signatures with \cPqb\ quarks.

This Letter presents the first measurement of the production cross section
of a \cPqb\ baryon, \PgLb, from fully reconstructed \psilam\ decays
in \pp\ collisions at $\sqrt{s}=7\TeV$
and complements the measurements of
\PBp~\cite{BPH-10-004}, \PBz~\cite{BPH-10-005}, and \PBzs~\cite{BPH-10-013}
production cross sections also performed by the Compact Muon Solenoid
(CMS) experiment at the LHC~\cite{JINST}.
The comparison of baryon production relative
to meson production resulting from the same initial \cPqb-quark momentum spectrum allows for tests
of differences in the hadronization process.
Such differences are particularly interesting in the context of heavy-baryon
production in relativistic heavy-ion collisions, where the medium could
significantly enhance the production of heavy baryons relative to
mesons~\cite{PhysRevLett.100.222301,Oh:2009zj,Ayala:2011pi}.
Furthermore, the \pp\ initial state at the LHC allows tests of baryon transport models,
which predict rapidity-dependent antibaryon/baryon asymmetries, in contrast to
baryon-antibaryon pair production, which typically results in equal
yields~\cite{Arakelyan:2007kq,Merino:2011sq}. Measurements
of the \PagLb\ to \PgLb\ cross-section ratio, \aPPratio, as functions
of \ptLb\ and \yLb\ allow for the first test of such models with heavy-quark baryons
at $\sqrt{s} = 7\TeV$.

Events with \PgLb\ baryons
reconstructed from their decays to the final state \psilam, with \psimumu\ and
\lamppi, are used to measure the differential cross sections
\dsdptbf, \dsdybf, and \aPPratio\ with respect
to the transverse momentum \ptLb\ and the rapidity \yLb, as well as the
integrated cross section times branching fraction for
\ptLb\ $> 10\GeV$ and \yLb\ $< 2.0$.
The cross section times
branching fraction is reported instead of the cross section itself because of the
$54\%$ uncertainty on \BLbpsilam~\cite{PDG2010}.
The cross section times branching fraction measurements are averaged over
particle and antiparticle states, while the ratio is computed by distinguishing the
two states via decays to \Pp\ or \Pap, respectively.

\section{Detector}
The data sample used in this analysis was collected by the CMS experiment
in
2011 and
corresponds to an integrated luminosity of
$1.86\pm0.04\invfb$~\cite{SMP-12-008}.
A detailed description of the detector may be found
elsewhere~\cite{JINST}. The main detector components used in this
analysis are the silicon tracker and the muon detection systems.

The silicon tracker measures charged particles within the pseudorapidity
range $|\eta| < 2.5$, where $\eta = -\ln[\tan{(\theta/2)}]$ and $\theta$ is the
polar angle of the track relative to the counterclockwise beam direction.
It consists of 1440 silicon pixel and 15\,148 silicon strip detector modules and
is located in the 3.8\unit{T} field of the superconducting solenoid. It provides an impact
parameter resolution of about 15\mum and a $\pt$
resolution of about 1.5\% for particles with transverse momenta up to
$100\GeV$. Muons are measured in the pseudorapidity range $|\eta|< 2.4$,
with detection planes made using three technologies: drift tubes, cathode strip chambers,
and resistive plate chambers.
Events are recorded with a two-level trigger system. The first level is composed of custom
hardware processors and uses information from the calorimeters and muon systems to select the most
interesting events.
The high-level trigger processor farm further decreases the event rate from
about 100\unit{kHz} to around 350\unit{Hz} before data storage.

\section{Event selection}
Early data taking conditions in 2011 utilized a loose dimuon trigger with the following requirements.
Events are selected requiring two oppositely charged muons
with dimuon transverse momentum greater than 6.9\GeV. Displaced muon pairs from
long-lived \cPqb-hadron decays are preferentially selected by further
requiring a transverse separation from the
mean \Pp\Pp\ collision position ("beamspot")
greater than three times its
uncertainty, where the uncertainty incorporates the vertex and beamspot measurements.
Also required at the trigger level are a dimuon vertex fit confidence level larger than
$0.5\%$ and $\cos \alpha > 0.9$,
where $\alpha$ is defined as the angle in the plane transverse to the beams between the dimuon
momentum and the
vector from the beamspot to the dimuon vertex.
The dimuon invariant mass
$m_{\Pgmp\Pgmm}$ is required to satisfy $2.9 < m_{\Pgmp\Pgmm} < 3.3\GeV$.
For the later $46\%$ of the dataset, the trigger was
tightened by increasing the dimuon vertex fit confidence level threshold to $10\%$ and imposing
kinematic requirements of $\pt^{\Pgm} > 3.5\GeV$ and $\left | \eta^{\Pgm}\right | < 2.2$ for
each of the muons.
The remaining 2011 data were recorded with even tighter triggers and are not used in the analysis.

Muon candidates are fully reconstructed by combining information from the
silicon tracker~\cite{TRK_10_001} and muon detectors, and are required to be within the
kinematic acceptance
region of $\pt^{\Pgm}>3.5\GeV$ and $\left | \eta^{\Pgm}\right | < 2.2$.
Muon candidates are further required to have a track $\chi^2$ per degree of freedom ${<}1.8$,
at least 11 silicon tracker hits, at least two hits in the pixel
system, and to be matched to at least one track segment in the muon system. Multiple
muon candidates are not allowed to share the same muon track
segments~\cite{MUO-10-002}.

Opposite-sign muon pairs are fit to a common vertex to form \JPsi\ candidates, which
are required to be within $150\MeV$ of the world-average \JPsi\ mass~\cite{PDG2010}.
The \JPsi\ candidates are also required to have $\pt$ greater than $7\GeV$, a dimuon vertex
fit confidence level larger than $0.5\%$, $\cos \alpha > 0.95$, and a transverse
separation of the vertex from the beamspot greater than three times its uncertainty.

The \PgL\ candidates are formed by fitting oppositely charged
tracks
to a common vertex. Each track
is required to have at least 6 hits in the silicon tracker, a
$\chi^2$ per degree of freedom ${<}5$, and a transverse impact parameter with respect to the beamspot
greater than 0.5 times its uncertainty. The proton candidate,
identified as the higher-momentum track, is required to have
$\pt > 1.0\GeV$. Misassignment of the correct proton track is found to be negligible from
simulation.
The reconstructed \PgL\ decay vertex
must have a $\chi^2$ per degree of freedom ${<}7$ and a transverse separation from the
beamspot at least five times larger than its uncertainty.
The invariant mass $m_{\Pp\Pgp}$ is required to be within $8\mev$ of the world-average
\PgL\ mass~\cite{PDG2010}. Candidates are rejected if $m_{\Pgpp\Pgpm}$ is within
$20\mev$ of the world-average \PKzS\ mass~\cite{PDG2010}.

\begin{figure}[!h]
  \begin{center}
   \includegraphics[width=0.49\textwidth]{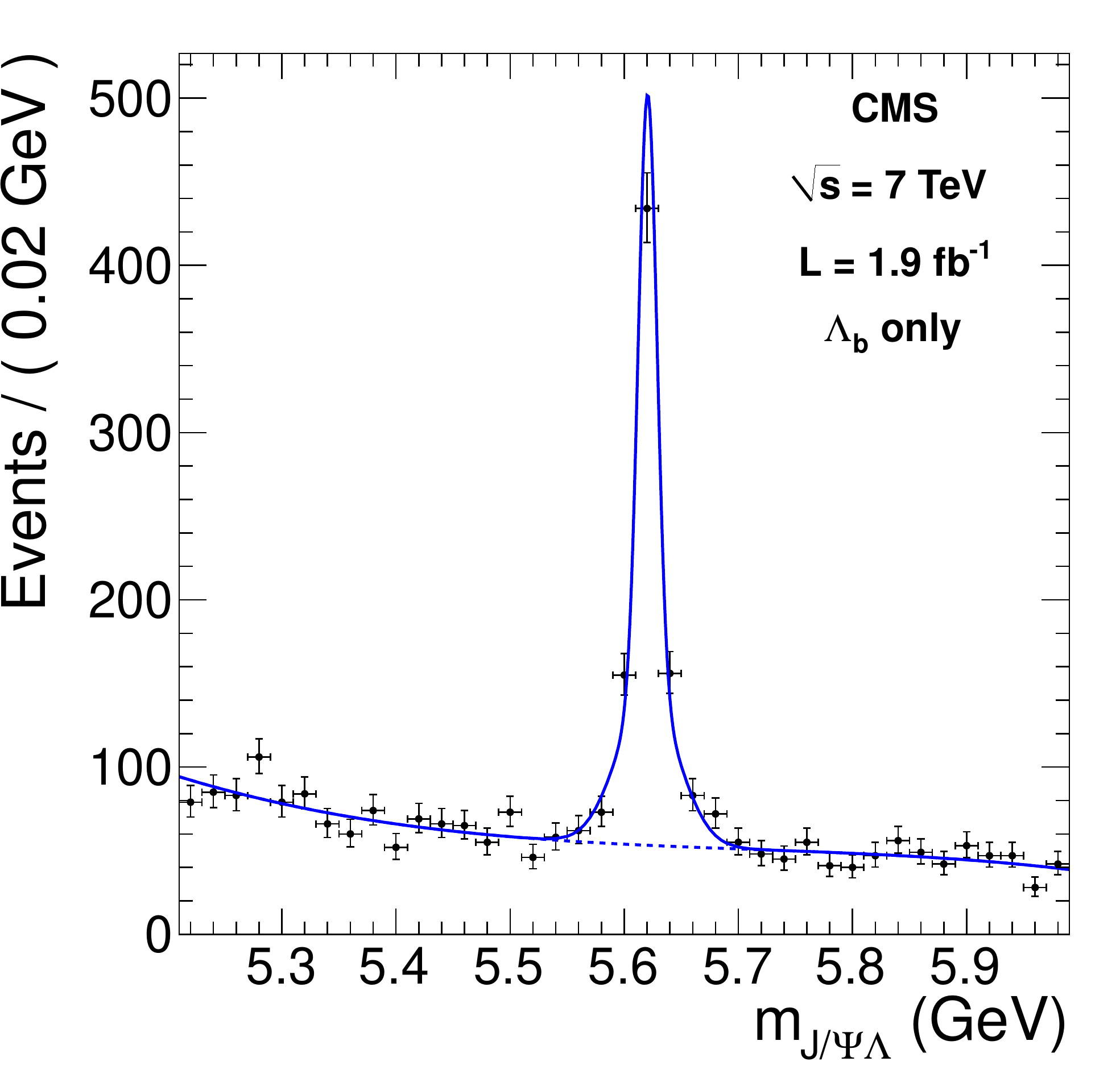}
   \includegraphics[width=0.49\textwidth]{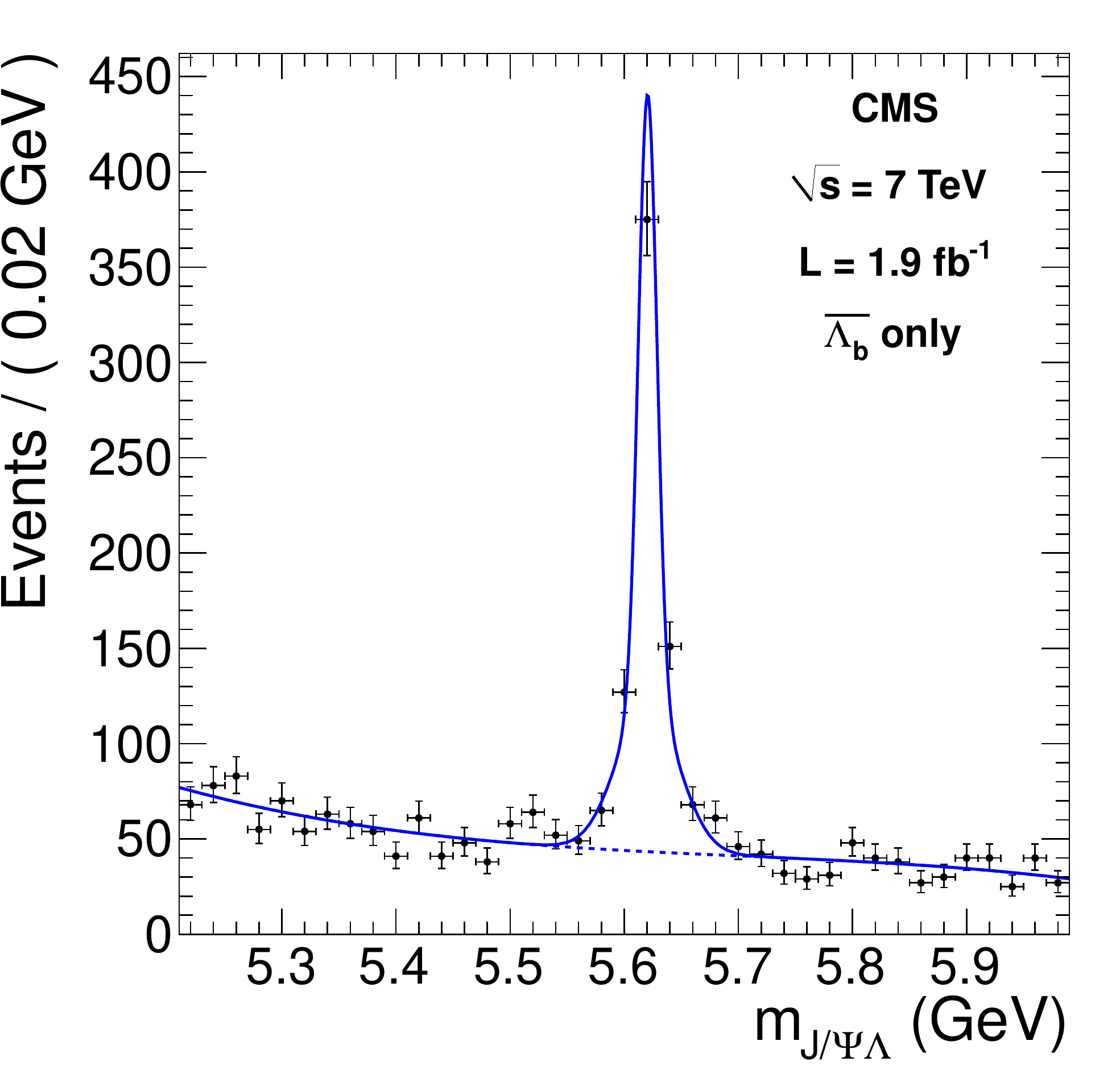}
    \caption{Fit results for the $\MLb$ distribution for \PgLb\ (\cmsLeft) and \PagLb\ (\cmsRight) for
    $\ptLb>10\GeV$ and $\yLb < 2.0$, where the dashed line shows the background fit function,
    the solid line shows the sum of signal and background, and the points indicate the data.}
    \label{fig:fit}
  \end{center}
\end{figure}

The \PgLb\ candidates are formed by combining a \JPsi\ candidate with a
\PgL\ candidate.
A vertex-constrained fit is performed with the two muons and the \PgL\ candidate, with the
invariant masses of the \JPsi\ and \PgL\ candidates constrained to their
world-average values~\cite{PDG2010}. The \PgLb\ vertex fit confidence level is
required to be greater than $1\%$ and the reconstructed $\PgLb$ mass must satisfy
$5.2<\MLb<6.0\gev$.
Multiple \PgLb\ candidates are found in less than $1\%$ of the events with at least one candidate
passing all selection criteria. In those cases, only the candidate with the highest
\PgLb\ vertex fit confidence level is retained. The \MLb\ distributions for selected
\PgLb\ and \PagLb\ candidates are shown in Fig.~\ref{fig:fit}.

\section{Efficiency determination}
The efficiency for triggering on and reconstructing \PgLb\ baryons
is computed with a combination of techniques using the data and
large samples of fully simulated Monte Carlo (MC) signal events generated with
\PYTHIA\ 6.422~\cite{PYTHIA}, decayed by \EVTGEN~\cite{EvtGen}, and simulated using
\GEANTfour~\cite{GEANT4}. The efficiency is factorized according to
\begin{equation}
\label{eq:eff}
\epsilon = \calA\cdot\epsilon^{\mu_1}_{\textrm{trig}}\cdot\epsilon^{\mu_2}_{\textrm{trig}}\cdot \epsilon^{\mu_1}_{\textrm{reco}}\cdot\epsilon^{\mu_2}_{\textrm{reco}}\cdot\epsilon^{\mu\mu}_{\textrm{trig}}\cdot\epsilon^{\PgLb}_{\textrm{sel}},
\end{equation}
where each term is described below.
The trigger ($\epsilon^{\mu_{\textrm{i}}}_{\textrm{trig}}$) and
muon-reconstruction efficiencies ($\epsilon^{\mu_\textrm{i}}_{\textrm{reco}}$) are obtained from a
large sample of inclusive
$\psimumu$ decays in data using a "tag-and-probe" technique similar to that described
in Ref.~\cite{BPH-10-002}, where one muon is identified with stringent quality requirements
and the second muon is identified using information either exclusively from the
tracker (to measure the trigger and offline muon-identification efficiencies) or from the
muon system (to measure the trigger and offline tracking efficiencies). While, in principle,
the inclusive $\psimumu$ sample can include signal events, which could bias the measurement, in
practice the fraction is negligibly small and provides an unbiased measurement of the muon efficiencies.

For the portion of the trigger efficiency that
depends on single-muon requirements ($\epsilon^{\mu_{\mathrm{i}}}_{\text{trig}}$),
the efficiency for a given \PgLb\ event is computed as the product
of the two single-muon efficiencies. However,
the trigger efficiencies for dimuon events where the muons bend toward each other
are up to $30\%$ lower than for events where the muons bend away from each other for certain
portions of the detector. This inefficiency arises when
the muon trajectories cross in the muon system, and one of the candidates is rejected
because of
shared hits. To account for this effect, the trigger
efficiencies for muons that bend toward and away from each
other are computed separately in data and the appropriate efficiency is applied to each class of
signal events.
This procedure naturally accounts for the correlations between the two single-muon efficiencies, as
confirmed in simulation.
The portions of the trigger efficiency that
depend on dimuon quantities ($\epsilon^{\mu\mu}_{\text{trig}}$) are measured from an inclusive \JPsi\ sample
collected with triggers where only single-muon requirements are applied.

The probabilities for the
muons to lie within the dimuon kinematic acceptance region ($\calA$) and for the \PgLb\ and
\PagLb\ candidates to pass the selection requirements ($\epsilon^{\PgLb}_{\text{sel}}$)
are determined from the simulated events.
To minimize the effect of the \PYTHIA\ modeling of the \ptLb\ and \yLb\ distributions
on the acceptance and efficiency calculations, the simulated events are reweighted to match the
kinematic distributions observed in the data.  The simulated events used for the
efficiency calculations have also been reweighted to match the measured distribution of the number
of \pp\ interactions per event (pileup). On average, there are six pileup interactions in the
data sample used in this analysis.
The efficiencies for hadron track
reconstruction~\cite{TRK-10-002}, \PgLb\ reconstruction~\cite{QCD-10-007},
and fulfilling the vertex quality requirements are found to be consistent between data and
simulation.

The total efficiency of this selection, defined as the fraction of $\Lbpsilam$ with $\psimumu$
and $\lamppi$ decays produced with $\ptLb>10\GeV$ and $\yLb<2.0$ that pass all criteria, is $0.73\%$. The efficiency
ranges from $0.3\%$ for $\ptLb$ 10--13$\GeV$ to $4.0\%$ for $\ptLb>28\GeV$, with the largest losses due to
the \PgL\ reconstruction (10--16$\%$ efficiency), the dimuon kinematic acceptance (12--63$\%$), and
the displaced dimuon trigger requirements (33--56$\%$). The efficiencies in bins of \ptLb\ and \yLb\ are
shown in Table~\ref{tab:results}.

To measure the ratio of antiparticle to particle cross sections \aPPratio, only the ratio of the
\PgLb\ and \PagLb\ detection efficiencies is needed. Many of
the efficiency contributions cancel in the ratio, including all the \JPsi\ and \Pgm\ efficiencies
since the particle and antiparticle states are indistinguishable. However, the \PgL\ and \PagL\
reconstruction
efficiencies differ because of different interaction cross sections with the detector material;
the \Pap\ are more likely to suffer a nuclear interaction and be lost, resulting in an efficiency that is
on average $13\%$ lower for \PagLb\ than for \PgLb, as
shown in Table~\ref{tab:ratio}.
The ratio of the
\PgLb\ and \PagLb\ selection efficiencies is calculated from simulation as described above for
the combined sample, where the simulation modeling of the detector interactions is validated by
comparing the number of hits reconstructed on tracks with that observed in data. The uncertainty on the
amount of
detector material and the appropriateness of simulated interaction cross sections are considered
as systematic uncertainties, as described in Section~\ref{sec:syst}.

\section{Fitting procedure}
The backgrounds are dominated by nonprompt $\JPsi$ production from \cPqb\ hadrons. The
dimuon invariant-mass distribution in data confirms that the contamination
from events containing a misidentified $\JPsi$ is negligible after all selection criteria have been
applied. Background events are distinguished from signal by their reconstructed
\MLb\ distribution, which is found to be in good agreement between data away from the signal peak and
simulated $\cPqb\to\JPsi X$ events. The \PgLb\ proper decay length distribution in data
confirms that the
background events arise from long-lived \cPqb\ hadrons, and therefore offers no additional
discriminating power between signal and background. The measured $m_{\Pp\Pgp}$ distribution
shows a purity of $77\%$ genuine \PgL\ events after applying the full selection criteria,
while the $m_{\Pgpp\Pgpm}$ distribution confirms
that more than $99.9\%$ of the \PKzS\ background is rejected by the kaon mass-window veto.

The $\PgLb$ yields are extracted from unbinned extended maximum-likelihood fits to the \MLb\ distribution
in bins of $\ptLb$ and $\yLb$ defined in Table~\ref{tab:results}. In each bin, the signal
is described by a double-Gaussian function with resolution parameters fixed to values found when fitting
simulated signal events and means set to a common value left free in the fit.
The background shape is modeled with a third-order polynomial, whose parameters are left
free to float independently in each bin. The ratio of antiparticle to particle yields is obtained
by simultaneously fitting the \PgLb\ and \PagLb\ mass distributions, with resolution parameters fixed
from the fit to the combined \PgLb\ and \PagLb\ simulated sample and
common mean allowed to float.
The background shapes are fit with separate third-order polynomials, whose parameters are left free in the fit.
The signal mass resolution varies as a function of \yLb, ranging from a mean of $11 \mev$ for central
\PgLb\ to $27 \mev$ for forward \PgLb\ events.

\section{Results}
\label{sec:results}

The fitted signal yields in each bin of $\ptLb$ and
$\yLb$ are summarized in Table~\ref{tab:results}.
Figure~\ref{fig:fit} shows the fits to the $\MLb$ distributions for \PgLb\ and
\PagLb\ candidates in the inclusive sample with $\ptLb>10\GeV$ and $\yLb < 2.0$.
The total number of signal events extracted from an inclusive fit is $1252\pm 42$,
where the uncertainty is statistical only.

\begin{table*}[!tb]
\topcaption{\PgLb\ $+$ \PagLb\ signal yield $n_{\rm sig}$,
efficiency $\epsilon$, and measured differential cross sections times branching fraction $\dsdptbf$
and $\dsdybf$, compared to the \POWHEG~\cite{POWHEG1,POWHEG2} and \PYTHIA~\cite{PYTHIA} predictions.
The uncertainties on
the signal yields are statistical only, while those on the efficiencies are systematic. The uncertainties
in the measured cross sections are statistical and systematic, respectively, excluding
the common luminosity
($2.2\%$)
and branching fraction ($1.3\%$) uncertainties.
The \POWHEG\ and \PYTHIA\ predictions also have uncertainties of $54\%$ due to \BLbpsilam,
which are not shown.
}
\label{tab:results}
\begin{center}
\begin{tabular*}{1.0\textwidth}{@{\extracolsep{\fill}}ccccccc}
\hline
\hline
$\ptLb$ \T \B      & $n_\text{sig}$ & $\epsilon$     & $\dsdptbf$ & $\POWHEG$ & $\PYTHIA$ \\
$(\!\GeV)$ \T \B      & events & $(\%)$     & $({\rm pb}/\!\GeV)$ & $({\rm pb}/\!\GeV)$ & $({\rm pb}/\!\GeV)$  \\
\hline
$10-13$   & $293 \pm 22 $ & $ 0.29 \pm 0.03 $   & $ 240 \pm 20 \pm 30 $	       & $110~^{+40}_{-30}$	 & 210  \\	
$13-15$   & $240 \pm 18 $ & $ 0.79 \pm 0.08 $   & $ 108 \pm 8 \pm 12 $	       & $54~^{+21}_{-12}$       & 102 \\	
$15-18$   & $265 \pm 19 $ & $ 1.54 \pm 0.16 $   & $  41 \pm 3 \pm 4 $	       & $29~^{+10}_{-6}$        & 55 \\	
$18-22$   & $207 \pm 16 $ & $ 2.34 \pm 0.23 $   & $  15.6 \pm 1.2 \pm 1.6 $    & $13.4~^{+4.5}_{-2.7}$   & 24.0 \\	
$22-28$   & $145 \pm 14 $ & $ 3.21 \pm 0.34 $   & $   5.3 \pm 0.5 \pm 0.6 $    & $5.3~^{+1.6}_{-1.1}$    & 9.3 \\
$28-50$   & $87 \pm 11 $  & $ 3.96 \pm 0.50 $   & $   0.70 \pm 0.09 \pm 0.09 $ & $0.89~^{+0.32}_{-0.15}$ & 1.42 \\
\sgline
$\yLb$ & $n_\text{sig}$ & $\epsilon$ & $\dsdybf$ & $\POWHEG$ & $\PYTHIA$ \\
       &  events        & $(\%)$     & $({\rm pb})$ &   $({\rm pb})$ & $({\rm pb})$  \\
\hline
$0.0-0.3$ & $233 \pm 17 $ & $0.74 \pm 0.09 $ & $ 370 \pm 30 \pm 50 $ & $180~^{+70}_{-40}$ & 330 \\
$0.3-0.6$ & $256 \pm 18 $ & $0.77 \pm 0.09 $ & $ 390 \pm 30 \pm 50 $ & $170~^{+60}_{-40}$ & 330 \\
$0.6-0.9$ & $206 \pm 16 $ & $0.81 \pm 0.09 $ & $ 300 \pm 20 \pm 30 $ & $170~^{+60}_{-40}$ & 320 \\
$0.9-1.2$ & $196 \pm 17 $ & $0.70 \pm 0.08 $ & $ 330 \pm 30 \pm 40 $ & $160~^{+60}_{-40}$ & 300 \\
$1.2-1.5$ & $189 \pm 17 $ & $0.67 \pm 0.09 $ & $ 330 \pm 30 \pm 50 $ & $150~^{+50}_{-40}$ & 280 \\
$1.5-2.0$ & $162 \pm 18 $ & $0.65 \pm 0.09 $ & $ 180 \pm 20 \pm 30 $ & $130~^{+50}_{-30}$ & 250 \\
\sgline
\hline
\end{tabular*}
\end{center}
\end{table*}

\begin{figure}[!h]
\begin{center}
\includegraphics[width=0.49\textwidth]{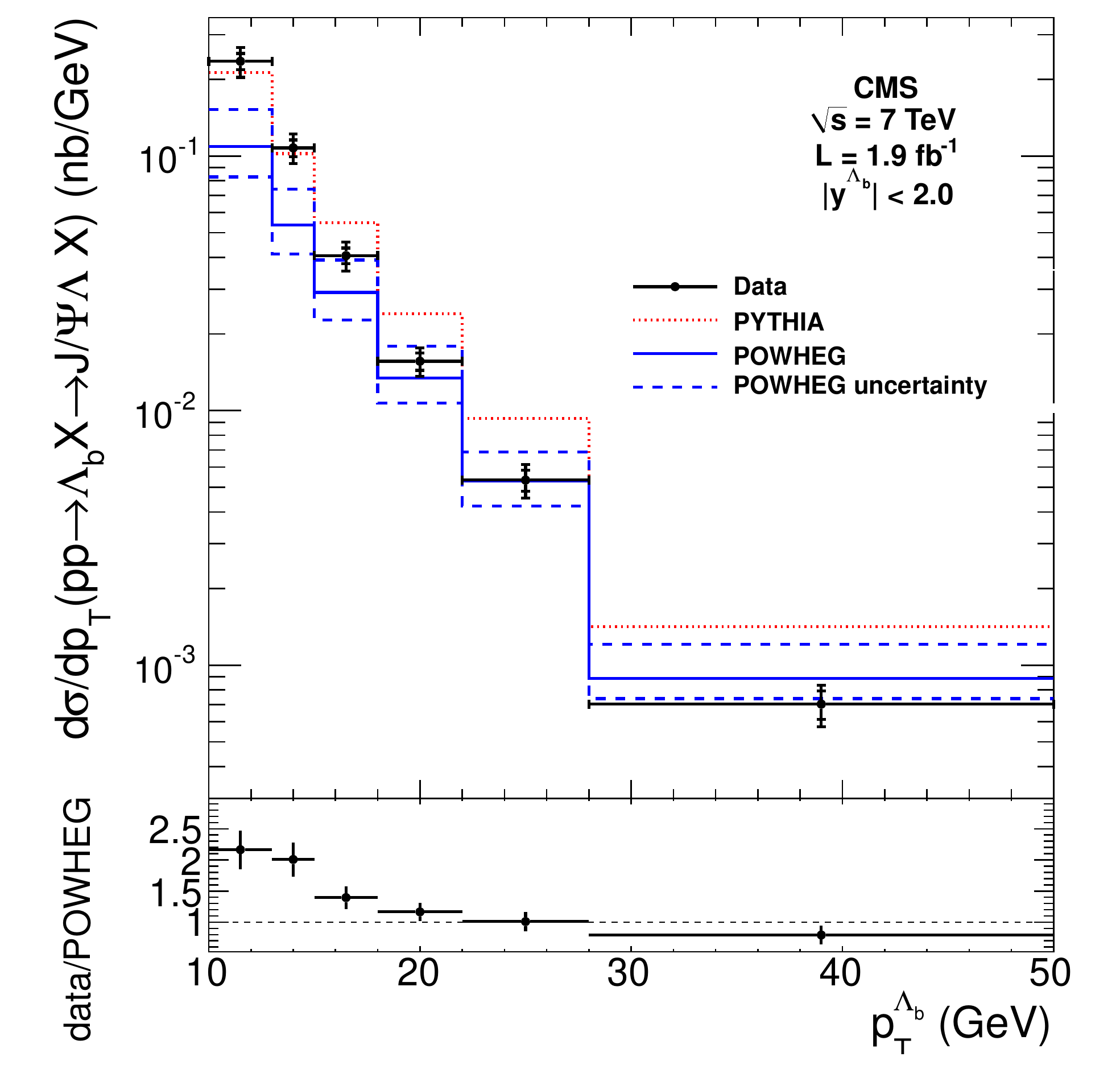}
\includegraphics[width=0.49\textwidth]{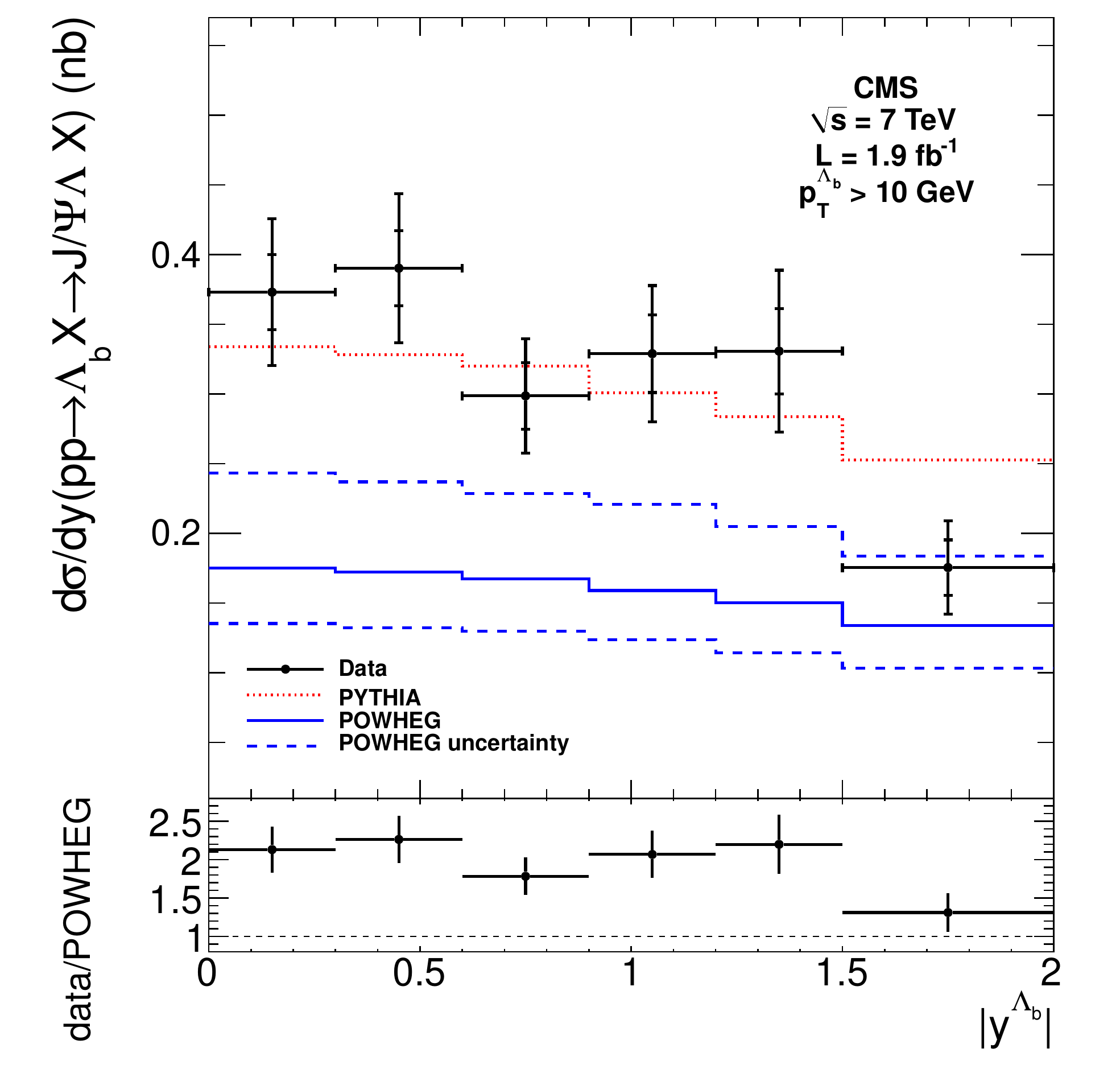}
\caption{Upper: Measured differential cross sections times branching fraction $\dsdptbf$ (\cmsLeft)
and $\dsdybf$ (\cmsRight) compared to the
theoretical predictions from \PYTHIA\ and \POWHEG.
The inner error bars correspond to the statistical uncertainties
and the outer ones represent the uncorrelated
systematic uncertainties added in quadrature to the statistical uncertainties.
The dashed lines show the uncertainties on the \POWHEG\ predictions.
Overall uncertainties of 2.2\% for the luminosity and 1.3\% for the \psimumu\ and
\lamppi\ branching fractions for the data are not shown, nor is the $54\%$ uncertainty
due to \BLbpsilam\ for the \PYTHIA\ and \POWHEG\ predictions. Lower: The ratio of the
measured values to the \POWHEG\ predictions. The error bars
include the statistical and uncorrelated systematic uncertainties on the data and the
shape-only uncertainties on the \POWHEG\ predictions.}
\label{fig:Xsec}
\end{center}
\end{figure}

The \PgLb\ differential cross section times branching fraction is calculated in bins of $\ptLb$ as
\begin{equation}
\frac{\rd\sigma(\Pp\Pp\to \PgLb X)}{\rd p_\mathrm{T}^{\PgLb}}\times\BLbpsilam = \frac{n_\text{sig}}
{2\cdot\epsilon \cdot\mathcal{B}\cdot\calL\cdot\Delta \ptLb},
\end{equation}
and similarly for $\yLb$, where $n_\text{sig}$ is the fitted number of signal
events in the given bin, $\epsilon$ is the average efficiency for signal $\PgLb$ and $\PagLb$ baryons
to pass all the selection criteria, \calL\ is the integrated luminosity,
$\Delta \ptLb$ is the bin size, and $\mathcal{B}$ is the product of branching
fractions $\mathcal{B}(\psimumu)=(5.93\pm 0.06)\times 10^{-2}$ and
$\mathcal{B}(\lamppi)= 0.639\pm 0.005$~\cite{PDG2010}.
The additional factor of two in the denominator accounts for our choice of quoting
the cross section for $\PgLb$ production only, while $n_{\rm sig}$ includes both \PgLb\ and
\PagLb.
The efficiencies are calculated separately for each bin, always considering only baryons produced with
$\yLb<2.0$ for \ptLb\ bins and $\ptLb>10\GeV$ for \yLb\ bins,
and taking into account bin-to-bin
migrations (0--2$\%$) because of the finite resolution on the measured \ptLb\ and \yLb. Equal production
of $\PgLb$ and $\PagLb$ is assumed for the efficiency, as predicted by \PYTHIA\ and as is consistent with
our measurement.

The measured differential cross sections times branching fraction versus $\ptLb$ and $\yLb$ are shown in
Fig.~\ref{fig:Xsec} and Table~\ref{tab:results}.  They are compared to predictions from the NLO
MC generator \POWHEG\ 1.0 with the \texttt{hvq} package~\cite{POWHEG1,POWHEG2} using a \cPqb-quark mass $m_b$ = $4.75\GeV$, renormalization
and factorization scales $\mu = \sqrt{m_{\cPqb}^2+\pt^2}$, CTEQ6M parton distribution
functions~\cite{CTEQ}, and \PYTHIA\ 6.422~\cite{PYTHIA} for the parton hadronization. The uncertainty on the
predicted cross section is calculated by varying the renormalization and factorization scales by
factors of two and, independently, $m_{\cPqb}$ by $\pm 0.25\gev$. The largest variation in each direction is
taken as the uncertainty. The data are also
compared to the \PYTHIA\ 6.422 prediction, using a \cPqb-quark mass of
$4.80\GeV$, CTEQ6L1 parton distribution functions, and the Z2 tune~\cite{Z2} to simulate the
underlying event. No attempt has been made to quantify the uncertainty on the \PYTHIA\ predictions.
The measured \pt\ spectrum falls faster than predicted by
\POWHEG\ and \PYTHIA, while the $|y|$ spectrum shape is in agreement with the
predictions within
uncertainties, as illustrated in the data-to-\POWHEG\ ratio plots shown in the lower panels of Fig.~\ref{fig:Xsec}.
The integrated cross section $\sigma(\pp\rightarrow\PgLb X)\times\BLbpsilam$ for
$\ptLb > 10\GeV$ and $\yLb < 2.0$,
calculated as the sum over all $\pt$ bins,
is \totalsigma, where the first uncertainty is statistical, and the
second is systematic. For the total cross section result, the highest \ptLb\ bin is fit without an upper bound
and has a yield of $97.0\pm 13.2$ events.  The total cross section measurement is in good agreement with the
prediction from \PYTHIA\ of $1.19\pm0.64$\unit{nb} and higher than the prediction from
\POWHEG\ of $0.63^{+0.41}_{-0.37}$\unit{nb}, where the uncertainties are dominated by
the $54\%$ uncertainty on \BLbpsilam~\cite{PDG2010}.

This result can be compared to previous CMS measurements of \PBp~\cite{BPH-10-004},
\PBz~\cite{BPH-10-005}, and \PBzs~\cite{BPH-10-013} production at $\sqrt{s} = 7\TeV$. To facilitate the comparison, the
\PBp\ and \PBz\ results are taken for the range $\ptb > 10\GeV$. Simulated events are generated with \MCATNLO~\cite{MCNLO}
with $m_b = 4.75\GeV$ and CTEQ6M parton distribution functions to determine the fraction of \PBp, \PBz, and \PBzs\ events
within the $\ptb$ and $\yb$ ranges used for their respective measurements with the $\pt > 10\GeV$ and
$|y| < 2.0$ requirements used in this analysis. Scaling by the appropriate ratio and using the world-average
values of $\BLbpsilam = (5.7 \pm 3.1)\times 10^{-4}$ and
$\calB(\PBzs\rightarrow\JPsi\phi) = (1.4 \pm 0.5)\times 10^{-3}$~\cite{PDG2010}, we determine the following
cross sections for $\ptb > 10\GeV$ and $\yb < 2.0$: $\sigma(\pp\rightarrow\PBp X) = 6.7\pm1.0\unit{$\mu$b}$;
$\sigma(\pp\rightarrow\PBz X) = 6.7\pm0.8\unit{$\mu$b}$; $\sigma(\pp\rightarrow\PBzs X) = 2.5\pm1.0\unit{$\mu$b}$
and $\sigma(\pp\rightarrow\PgLb X) = 2.1\pm1.1\unit{$\mu$b}$,
where the uncertainties are the quadrature sum of the statistical and systematic components. No uncertainty has
been included for the phase-space extrapolation based on \MCATNLO~\cite{MCNLO}. The large systematic uncertainties
for $\sigma(\pp\rightarrow\PBzs X)$ and $\sigma(\pp\rightarrow\PgLb X)$ are dominated by the poorly known
branching fractions $\BLbpsilam$ and $\calB(\PBzs\rightarrow\JPsi\phi)$, respectively.
The ratios among the four results are in good agreement with the world-average \cPqb-quark
fragmentation results~\cite{PDG2010}.

The world-average \cPqb-quark fragmentation results assume that the fractions
are the same for \cPqb\ jets originating from $Z$ decays at LEP and directly from $\Pp\Pap$ collisions at the
Tevatron.
However, measurements of $f_{\PgLb}$ performed at LEP~\cite{Abreu:1995me,Buskulic:1996sm}
and at the Tevatron~\cite{PhysRevD.77.072003}
show discrepancies. A recent result~\cite{Aaij:2011jp} from the
LHCb Collaboration measures a strong \pt\ dependence of the ratio of \PgLb\ production to \PB-meson production,
$f_{\PgLb}/(f_u+f_d)$, with $f_{\PgLb}\equiv\calB(b\rightarrow\PgLb)$ and $f_q\equiv\calB(b\rightarrow\PB_q)$.
Larger $f_{\PgLb}$ values are observed at lower \pt,
which suggests that the discrepancy observed between the LEP and Tevatron data
may be due to the lower \pt\ of the \PgLb\ baryons produced at the Tevatron.

\begin{figure}[!h]
\begin{center}
\includegraphics[width=0.48\textwidth]{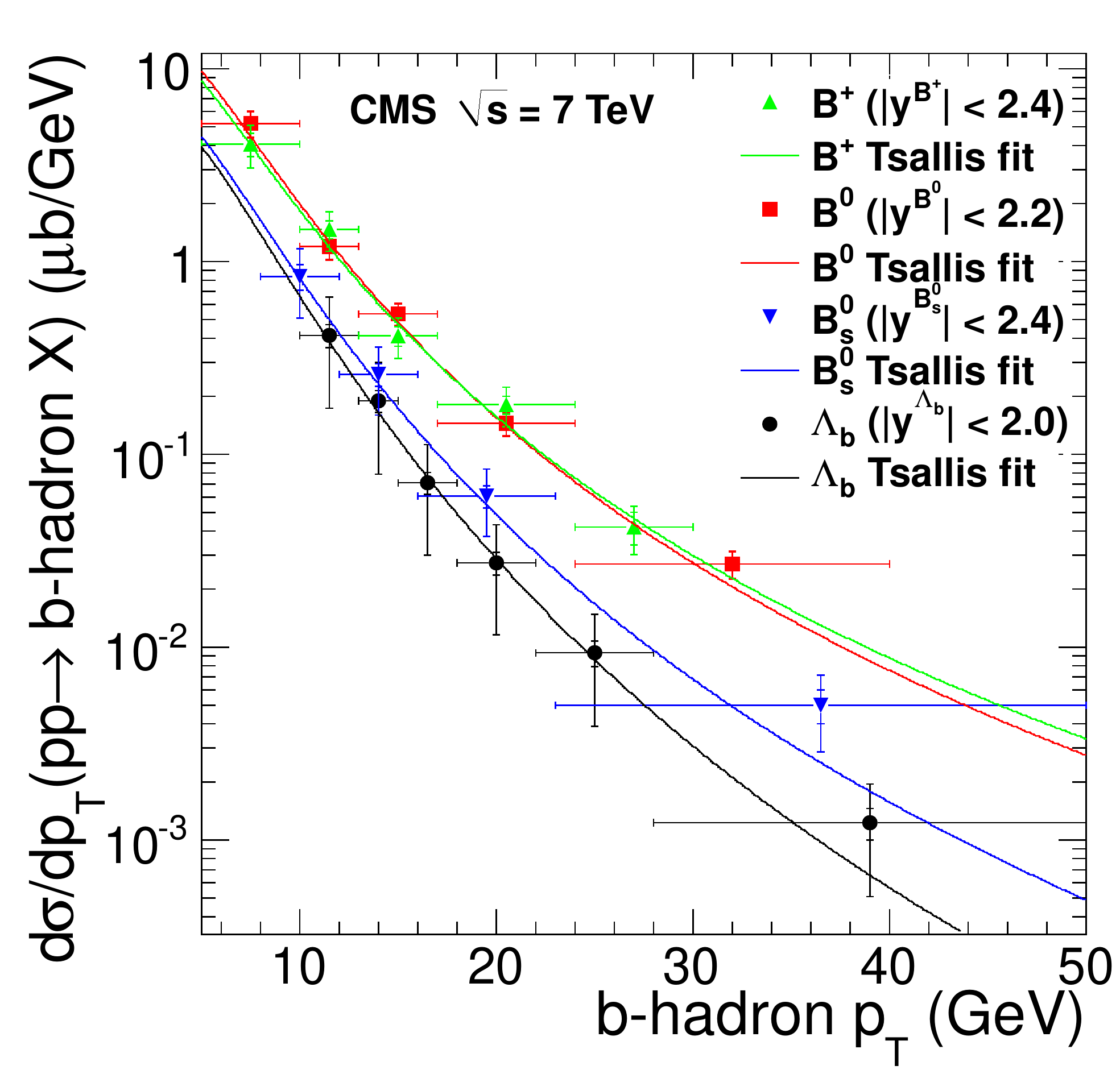}
\includegraphics[width=0.48\textwidth]{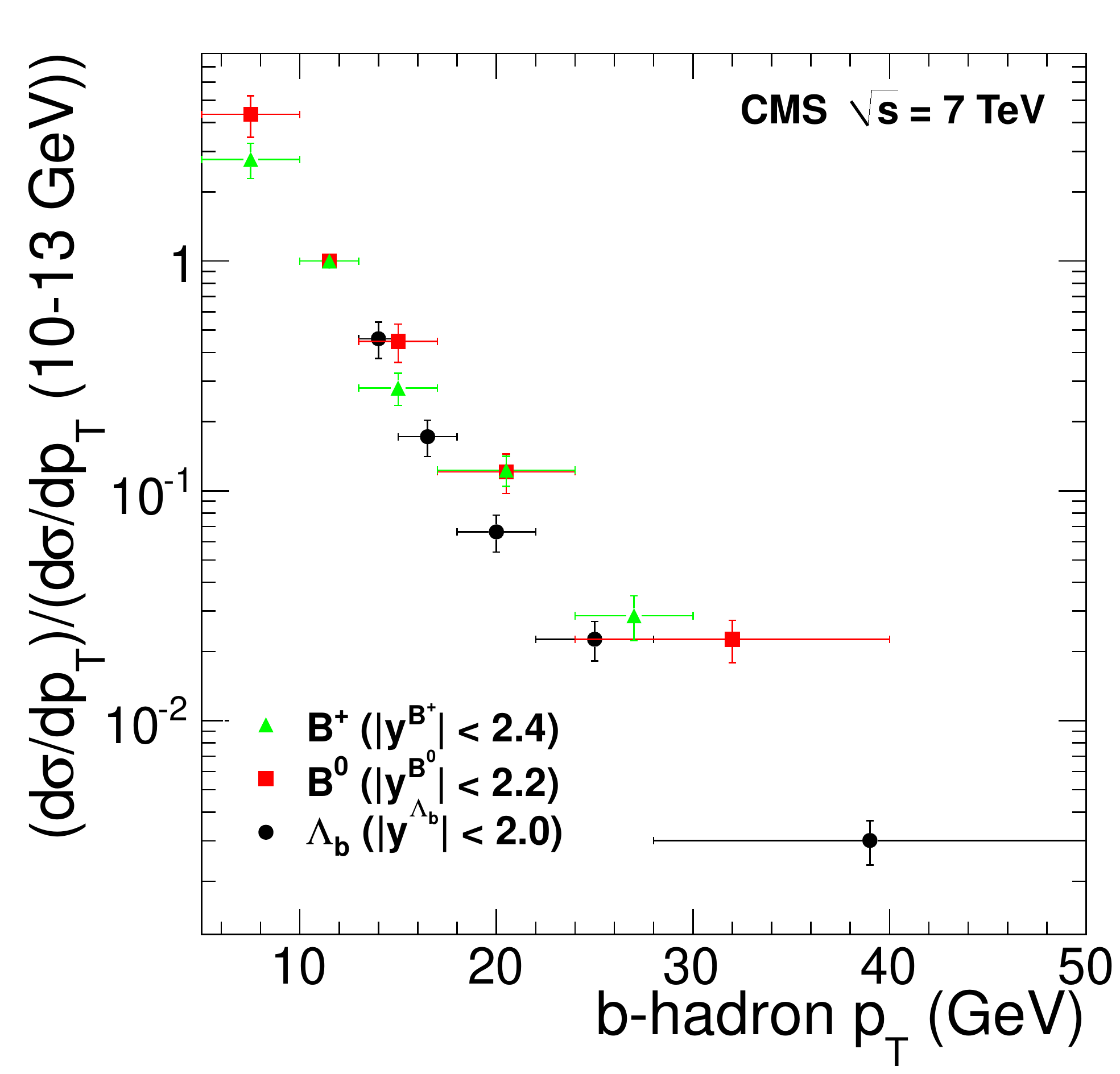}
\caption{Comparison of production rates for \PBp~\cite{BPH-10-004}, \PBz~\cite{BPH-10-005},
\PBzs~\cite{BPH-10-013}, and \PgLb\ versus \pt. The \cmsLeft\ plot shows the absolute comparison, where the
inner error bars correspond to the total bin-to-bin uncertainties, while the
outer error bars represent the total bin-to-bin and normalization uncertainties added in quadrature. Fits to the
Tsallis function~\cite{Tsallis} for each distribution are also shown.The overall
uncertainties for \PBzs\ and \PgLb\ are dominated by large uncertainties on $\calB(\PBzs\rightarrow\JPsi\phi)$ and
\BLbpsilam, respectively. The \cmsRight\ plot
shows a shape-only comparison where the data are normalized to the 10$-$13 $\gev$ bin in \pt\ and the
error bars show the bin-to-bin uncertainties only. \PBzs\ is omitted because the 10$-$13 $\gev$ bin is
not available for the common normalization.}
\label{fig:comparison}
\end{center}
\end{figure}

A comparison of this and previous CMS results for \cPqb-hadron production versus
\pt\ is shown in the left plot of Fig.~\ref{fig:comparison}, where the data are fit to the Tsallis
function~\cite{Tsallis},
\begin{equation}
\frac{1}{N} \,\frac{\rd N}{\rd\pt} = C\, \pt\left[1+ \frac{\sqrt{p_\mathrm{T}^2+m^2}-m}{nT}\right]^{-n}.
\end{equation}
Here $C$ is a normalization parameter, $T$ and $n$ are shape parameters, $m$ is the mass of the \cPqb\ hadron
and $N$ is the \cPqb-hadron yield. The statistical and bin-to-bin
systematic uncertainties are used in the fits. The $T$ parameter represents the
inverse slope parameter of an exponential, which dominates at low $\pt$.
Since our data do not constrain that region well, $T$ is fixed to the mean
value found from fitting the \PBp\ and \PBz\ distributions, where the \pt\ threshold
is lowest. The result of $T = 1.10$\GeV\ is used to obtain the following values of the $n$
parameter, which controls the power-law behavior at high $\pt$: $n(\PBp) = 5.5 \pm 0.3$, $n(\PBz) = 5.8 \pm 0.3$,
$n(\PBzs) = 6.6\pm0.4$, and $n(\PgLb) = 7.6\pm0.4$. The larger $n$ value for \PgLb\ indicates a
more steeply falling \pt\ distribution than observed for the mesons, also suggesting that the
production of \PgLb\ baryons, relative to \PB\ mesons, varies as a function of \pt, with a
larger \PgLb/\PB\ ratio at lower transverse momentum. The right plot of Fig.~\ref{fig:comparison} shows the \ptLb\ spectrum
shape compared to \PBp\ and \PBz, where the distributions are normalized to the common bin with \pt\ = 10$-$13 $\gev$.

\begin{figure}[!h]
\begin{center}
\includegraphics[width=0.49\textwidth]{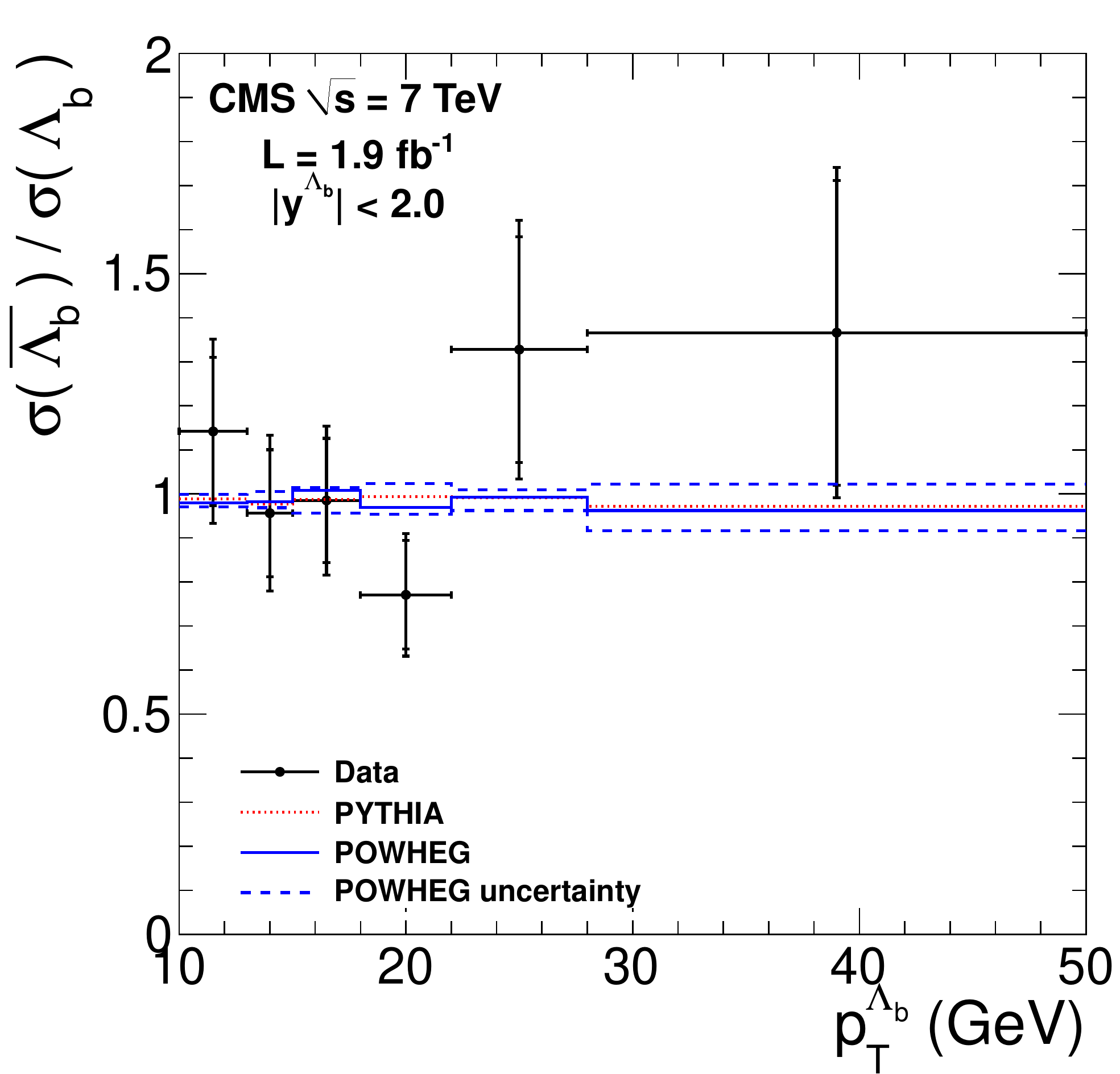}
\includegraphics[width=0.49\textwidth]{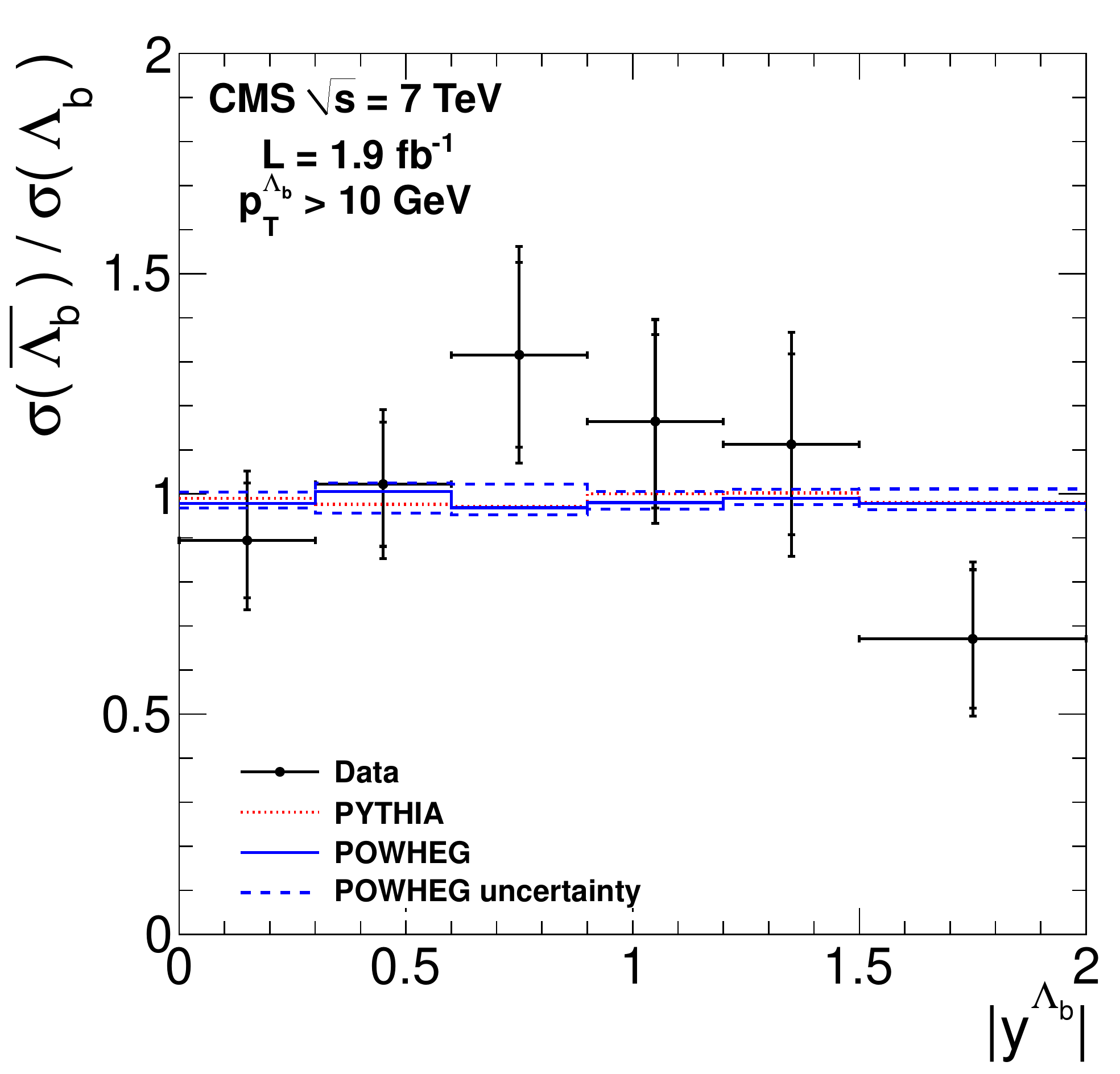}
\caption{Measured \aPPratio\ (points) versus
$\ptLb$ (\cmsLeft) and $\yLb$ (\cmsRight), compared to the
theoretical predictions from \PYTHIA\ (red dashed line) and \POWHEG\ (blue solid line).
The inner error bars correspond to the statistical uncertainties,
and the outer error bars represent the uncorrelated
systematic uncertainties added in quadrature to the statistical uncertainties. The dashed blue
lines show the uncertainties of the \POWHEG\ predictions.}
\label{fig:Asym}
\end{center}
\end{figure}

The ratio \aPPratio\ is calculated in bins of $\ptLb$ or $\yLb$ as
\begin{equation}
\aPPratio = \frac{n_\text{sig}^{\PagLb}}{n_\text{sig}^{\PgLb}}\times\frac{\epsilon (\PgLb)}{\epsilon (\PagLb)},
\end{equation}
where $n_\text{sig}^{\PagLb}$ and $n_\text{sig}^{\PgLb}$ are the antiparticle and particle yields in a given bin,
and $\epsilon (\PgLb)$ and $\epsilon (\PagLb)$ are the particle and antiparticle efficiencies for a given bin,
always considering only baryons produced with $\yLb<2.0$ for \ptLb\ bins and $\ptLb>10\GeV$ for \yLb\ bins.
The results versus \ptLb\ and \yLb\ are shown in Fig.~\ref{fig:Asym} and Table~\ref{tab:ratio}.
The ratio \aPPratio\ is found to be consistent with unity and constant as a function of both \ptLb\ and \yLb,
within the uncertainties, as predicted by \POWHEG\ and \PYTHIA.
Therefore, no evidence of
increased baryon production at forward pseudorapidities is observed within the available statistical precision
for the kinematic regime investigated. The integrated \aPPratio\ for
$\ptLb>10\GeV$ and $\yLb<2.0$ is \totalasym, where
the first uncertainty is statistical and the second is systematic.

\begin{table*}[!htbp]
\topcaption{Uncorrected signal yield ratio $n_\text{sig}^{\PagLb}$/$n_\text{sig}^{\PgLb}$, efficiency ratio $\epsilon(\PagLb)/\epsilon(\PgLb)$, and
efficiency-corrected ratio \aPPratio, compared to the \POWHEG~\cite{POWHEG1,POWHEG2} and \PYTHIA~\cite{PYTHIA}
predictions for the corrected
ratio. The uncertainties in the corrected ratio are statistical and systematic, respectively. The uncertainties on
the uncorrected yield ratio are statistical only and on the efficiency ratio are systematic only.}
\label{tab:ratio}
\begin{center}
\begin{tabular}{lccccc}
\dbline
                          & Uncorrected &                                          & Data & \POWHEG & \PYTHIA \\
$\ptLb~(\GeVns)$ \T \B    & $n_\text{sig}^{\PagLb}$/$n_\text{sig}^{\PgLb}$ &  $\epsilon(\PagLb)$/$\epsilon(\PgLb)$ & \aPPratio & \aPPratio & \aPPratio \\

\sgline
10--13  &  0.96$\pm$0.14 & 0.84$\pm$0.09 & 1.14$\pm$0.17$\pm$0.12 & $0.98~^{+0.02}_{-0.01}$ & 0.99 \\
13--15  &  0.76$\pm$0.11 & 0.79$\pm$0.09 & 0.96$\pm$0.14$\pm$0.10 & $0.98~^{+0.02}_{-0.01}$ & 0.98 \\
15--18  &  0.89$\pm$0.13 & 0.90$\pm$0.09 & 0.98$\pm$0.14$\pm$0.09 & $1.01~^{+0.01}_{-0.05}$ & 0.99 \\
18--22  &  0.73$\pm$0.12 & 0.95$\pm$0.08 & 0.77$\pm$0.12$\pm$0.07 & $0.97~^{+0.05}_{-0.02}$ & 0.99 \\
22--28  &  1.26$\pm$0.24 & 0.94$\pm$0.10 & 1.33$\pm$0.26$\pm$0.14 & $0.99~^{+0.02}_{-0.03}$ & 0.99 \\
28--50  &  0.99$\pm$0.25 & 0.72$\pm$0.08 & 1.37$\pm$0.35$\pm$0.14 & $0.96~^{+0.06}_{-0.04}$ & 0.97 \\
\sgline
           & Uncorrected   &                                       & Data & \POWHEG & \PYTHIA \\
$\yLb$   &  $n_\text{sig}^{\PagLb}$/$n_\text{sig}^{\PgLb}$ &  $\epsilon(\PagLb)$/$\epsilon(\PgLb)$ & \aPPratio & \aPPratio & \aPPratio \\
\sgline
0.0--0.3  & 0.71$\pm$0.10 & 0.79$\pm$0.08 & 0.89$\pm$0.13$\pm$0.09 & $0.98~^{+0.02}_{-0.01}$ & 0.99 \\
0.3--0.6  & 0.92$\pm$0.13 & 0.90$\pm$0.08 & 1.02$\pm$0.14$\pm$0.09 & $1.01~^{+0.01}_{-0.05}$ & 0.98 \\
0.6--0.9  & 1.16$\pm$0.18 & 0.88$\pm$0.09 & 1.32$\pm$0.21$\pm$0.13 & $0.97~^{+0.05}_{-0.02}$ & 0.97 \\
0.9--1.2  & 0.99$\pm$0.17 & 0.85$\pm$0.09 & 1.16$\pm$0.20$\pm$0.12 & $0.98~^{+0.03}_{-0.02}$ & 1.00 \\
1.2--1.5  & 0.92$\pm$0.17 & 0.82$\pm$0.11 & 1.11$\pm$0.20$\pm$0.15 & $0.99~^{+0.02}_{-0.01}$ & 1.00 \\
1.5--2.0  & 0.66$\pm$0.16 & 0.99$\pm$0.11 & 0.67$\pm$0.16$\pm$0.08 & $0.98~^{+0.03}_{-0.02}$ & 0.98 \\
\dbline
\end{tabular}
\end{center}
\end{table*}

\section{Systematic uncertainties}
\label{sec:syst}
The cross section is affected by systematic uncertainties on the signal yields and efficiencies that are uncorrelated
bin-to-bin and can affect the shapes of the distributions, and by the uncertainties on branching fractions
and integrated luminosity, which are common to all bins and only affect the overall normalization.
The uncertainties on the signal yields arise from the following sources:
\begin{itemize}
\item Signal shape uncertainty ($1$--$6\%$): evaluated from the variations when floating the means of the two Gaussians (set to a common value)
in data or by using a single Gaussian shape.
\item Background shape uncertainty ($1$--$2\%$): evaluated from the variation when using a second-order polynomial, exponential,
or third-order polynomial fit in the restricted range 5.4--6.0\GeV.
\item Final-state radiation ($0$--$1\%$): evaluated by removing it from the simulation and taking half of the difference in the results.
\end{itemize}
The uncertainties on the efficiencies arise from the following sources:
\begin{itemize}
\item Pion/proton/$\Lambda$ reconstruction efficiency uncertainty ($8\%$): evaluated by varying the simulated
detector material~\cite{CMS-NOTE-2010-010},
alignment, and beamspot position, and by varying the reconstruction cuts, by using different event simulations, and comparing the
measured \PgL\ lifetime~\cite{QCD-10-007}, which is sensitive to the efficiency correction, to the world-average value~\cite{PDG2010}.
\item Tag-and-probe statistical uncertainties ($4$--$6\%$): evaluated by propagating statistical
uncertainties from the data-driven determination of the single-muon efficiencies.
\item Tag-and-probe systematic uncertainties ($1$--$7\%$): evaluated as the difference between the true efficiency in simulation and
the efficiency calculated with the tag-and-probe procedure applied to simulated events.
\item Statistical precision of the simulated event samples ($3$--$4\%$): calculated for the dimuon acceptance and reconstruction efficiencies.
\item Simulation modeling of the $\PgLb$ kinematic distributions ($0$--$5\%$): evaluated as half of the difference
due to the kinematic reweighting.
\item \GEANTfour\ \Pap\ cross section ($1$--$4\%$):
evaluated by considering an alternative cross section model in \GEANTfour\ (CHIPS)~\cite{GEANT4} for
\Pap\ cross sections for interacting with material in the detector~\cite{Alice_ppbar}
and taking the difference in the efficiency as a systematic uncertainty.
\item Unknown $\PgLb$ polarization ($1$--$4\%$): evaluated by generating samples of events with the \PgLb\ spin fully aligned or anti-aligned
with the normal to the plane defined by the \PgLb\ momentum and the \pp\ beam direction in the laboratory frame
and taking the average difference in the efficiency when compared to the nominal
analysis, which is performed with unpolarized simulated events.
\item Pileup ($0$--$4\%$): evaluated by varying the number of pileup interactions in simulated events by the uncertainty of the measured pileup
interaction distribution.
\item Muon kinematics ($0$--$2\%$): evaluated as the difference in the simulated efficiency when reweighting the muon \pt\ to match the distribution
measured with muons from the inclusive \JPsi\ sample used in the tag-and-probe measurements.
\item Effect of events migrating between \pt\ and $y$ bins due to resolution ($0$--$1\%$): evaluated as half of the correction deduced from
simulated events.
\end{itemize}

The bin-to-bin systematic uncertainty
is computed as the sum in quadrature of the individual uncertainties and is
summarized in Table~\ref{tab:results}.  In addition, there are normalization
uncertainties of $2.2\%$ from the luminosity measurement~\cite{SMP-12-008}
and of $1.3\%$ from the \psimumu\ and \lamppi\ branching fractions~\cite{PDG2010}.
For the total cross section result computed from the sum of \pt\ bins, only the
signal and background shapes, and the tag-and-probe and simulation statistical uncertainties
are treated as uncorrelated. As bin-to-bin correlations cannot be ruled out for the
remaining sources of systematic uncertainty, the contribution in each \pt\ bin is added
linearly to compute the sum to ensure that the uncertainty is not underestimated.

Many of these systematic effects cancel in the \aPPratio\ ratio measurement. The remaining uncertainties are
from the signal shape ($2$--$8\%$), background shape ($1$--$3\%$),
\GEANTfour\ \Pap\ cross section ($1$--$7\%$), variation of detector material ($5\%$), and
statistical precision of the simulated samples ($6$--$8\%$), which are evaluated as described above.
The total systematic uncertainty is computed
as the quadrature sum of the individual uncertainties and is summarized in Table~\ref{tab:ratio}.

\section{Conclusions}
In summary, the first measurements of the differential cross sections times branching fraction
$\dsdptbf$ and $\dsdybf$ for
$\PgLb$ baryons produced in $\Pp\Pp$ collisions at $\sqrt{s}=7\TeV$ have been presented.
The measurements are given for $\ptLb > 10\GeV$ and $\yLb < 2.0$.
The \ptLb\ distribution falls faster than both the measured \pt\ spectra from \cPqb\ mesons and
the predicted spectra from the NLO MC \POWHEG\ and the leading-order MC \PYTHIA. The total cross
section and rapidity distribution are consistent with both predictions within large
uncertainties. The measured \aPPratio\ ratio
is consistent with unity and constant as a function of both
\ptLb\ and \yLb.

\section*{Acknowledgments}

We congratulate our colleagues in the CERN accelerator departments for the excellent performance of the LHC machine.
We thank the technical and administrative staff at CERN and other CMS institutes, and acknowledge support from: FMSR
(Austria); FNRS and FWO (Belgium); CNPq, CAPES, FAPERJ, and FAPESP (Brazil); MES (Bulgaria); CERN; CAS, MoST, and
NSFC (China); COLCIENCIAS (Colombia); MSES (Croatia); RPF (Cyprus); MoER, SF0690030s09 and ERDF (Estonia); Academy
of Finland, MEC, and HIP (Finland); CEA and CNRS/IN2P3 (France); BMBF, DFG, and HGF (Germany); GSRT (Greece); OTKA
and NKTH (Hungary); DAE and DST (India); IPM (Iran); SFI (Ireland); INFN (Italy); NRF and WCU (Korea); LAS
(Lithuania); CINVESTAV, CONACYT, SEP, and UASLP-FAI (Mexico); MSI (New Zealand); PAEC (Pakistan); MSHE and NSC
(Poland); FCT (Portugal); JINR (Armenia, Belarus, Georgia, Ukraine, Uzbekistan); MON, RosAtom, RAS and RFBR
(Russia); MSTD (Serbia); MICINN and CPAN (Spain); Swiss Funding Agencies (Switzerland); NSC (Taipei); TUBITAK and
TAEK (Turkey); STFC (United Kingdom); DOE and NSF (USA).
Individuals have received support from the Marie-Curie programme and the European Research Council (European Union);
the Leventis Foundation; the A. P. Sloan Foundation; the Alexander von Humboldt Foundation; the Belgian Federal
Science Policy Office; the Fonds pour la Formation \`a la Recherche dans l'Industrie et dans l'Agriculture
(FRIA-Belgium); the Agentschap voor Innovatie door Wetenschap en Technologie (IWT-Belgium); the Council of Science
and Industrial Research, India; and the HOMING PLUS programme of Foundation for Polish Science, cofinanced from
European Union, Regional Development Fund.

\bibliography{auto_generated}   

\cleardoublepage \appendix\section{The CMS Collaboration \label{app:collab}}\begin{sloppypar}\hyphenpenalty=5000\widowpenalty=500\clubpenalty=5000\textbf{Yerevan Physics Institute,  Yerevan,  Armenia}\\*[0pt]
S.~Chatrchyan, V.~Khachatryan, A.M.~Sirunyan, A.~Tumasyan
\vskip\cmsinstskip
\textbf{Institut f\"{u}r Hochenergiephysik der OeAW,  Wien,  Austria}\\*[0pt]
W.~Adam, T.~Bergauer, M.~Dragicevic, J.~Er\"{o}, C.~Fabjan, M.~Friedl, R.~Fr\"{u}hwirth, V.M.~Ghete, J.~Hammer, N.~H\"{o}rmann, J.~Hrubec, M.~Jeitler, W.~Kiesenhofer, V.~Kn\"{u}nz, M.~Krammer, D.~Liko, I.~Mikulec, M.~Pernicka$^{\textrm{\dag}}$, B.~Rahbaran, C.~Rohringer, H.~Rohringer, R.~Sch\"{o}fbeck, J.~Strauss, A.~Taurok, P.~Wagner, W.~Waltenberger, G.~Walzel, E.~Widl, C.-E.~Wulz
% ...
\vskip\cmsinstskip
\dag:~Deceased\\
% ...

\end{sloppypar}
\end{document}